# Detecting onset of chain scission and crosslinking of $\gamma$-ray irradiated elastomer surfaces using frictional force microscopy

S. Banerjee[a*], N. K. Sinha[b], N. Gayathri[a],

D. Ponraju[c], S. Dash[a], A. K. Tyagi[a] and Baldev Raj

[a] *Materials Science Division*

[b] *Innovative Design Engineering and Synthesis Section*

[c] *Radiological Safety Division*

*Indira Gandhi Centre for Atomic Research,*

*Kalpakkam   603102, T.N, India*

## Abstract

We report here that atomic force microscope (AFM) in frictional force mode can be used to detect onset of chain scission and crosslinking in polymeric and macromolecular samples upon irradiation. A systematic investigation to detect chain scission and crosslinking of two elastomers: (1) Ethylene-propylene-diene monomer rubber (EPDM) and (2) Fluorocarbon rubber (FKM) upon $\gamma$-ray irradiation has been carried out using frictional force microscopy (FFM). From the AFM results we observed that both the elastomers show a systematic smoothening of its surfaces, as the $\gamma$-ray dose rate increases. However, the frictional property studied using FFM of the sample surfaces show an initial increase and then a decrease as a function of dose rate. This behavior of increase in its frictional property has been attributed to the onset of chain scission and the subsequent decrease in friction has been attributed to the onset of crosslinking of the polymer chains. The evaluated qualitative and semi-quantitative changes observed in the overall frictional property as a function of $\gamma$-ray dose rate for the two elastomers are presented in this paper.

---
[*] Corresponding author and on Deputation from Surface Physics Division, Saha Insitute of Nuclear Physics, 1/AF Bidhannagar, Kolkata, Email:sangam@hotmail.com

## I. INTRODUCTION

Crosslinking of polymer chains improves the physical properties of rubber[1,2]. Crosslinking is the key to the elastic, or "rubbery" nature of any elastomer. On crosslinking, the glass transition temperature ($T_g$)[3,4], elasticity[5,6] and tensile strength[5–7] of the rubber increases and its elongation[5,6,8], wear and friction decreases[9]. These properties warrant safety and reliability of the rubber. Crosslinking is the process of forming a three-dimensional network structure from a linear polymer by a chemical or a physical method. The crosslinking in rubbers can be achieved by (1) conventional thermal curing using chemical reagents (first-generation technology), and (2) physically inducing crosslinking by high-energy radiation (second-generation technology). The second method is gaining importance because of certain advantages. Radiation generates radicals and ions in the medium without any use of additional chemical catalyst and hence the electrical property of the rubber can be almost preserved and can be carried out at controlled temperature. Radiation induced crosslinking is a physically induced chemical reaction which is easier and preferable for continuous curing of the rubber. Crosslinking reaction pathways and mechanisms are still not completely understood and whether the reaction occurs mainly via radical or an ionic pathway or both is still one of the fundamental issues. But, it is believed that upon irradiation by high energy radiation such as x-ray, proton, electron, neutron or $\gamma$-ray, high local concentrations of free radicals are formed in the rubber molecules.

Radiation causes chain scission of the polymer causing production of excited macromolecular radicals. They polymerise, combine or crosslink only beyond certain critical concentration of free radical generation. The physical property of the irradiated rubber is determined by the competition between the rate of chain scission and crosslinking i.e., on the ratio of crosslinking to chain scission. Numerous studies related to the modification of physical properties upon high energy radiation have been carried out[10–19]. In this paper we report systematic study on the evolution of local frictional property of elastomers upon $\gamma$-ray irradiation. The local frictional property has been studied using atomic force microscope (AFM) in a frictional force mode (frictional force microscopy - FFM). To our knowledge no study to detect onset of chain scission and crosslinking on elastomers upon $\gamma$-ray irradiation using FFM has been reported. We would like to show in the present study that FFM can be used to detect the competition between the chain scission and crosslinking of the polymer upon high energy irradiation.

## II. EXPERIMENTAL

Two sets of samples (1) Ethylene-propylene-diene monomer rubber (EPDM) with the chemical formula $[-CH_2-CH_2-CH_2-CH(CH_3)-]_n$ and (2) Fluorocarbon rubber (terpolymer of Vinilidenefluoride, hexafluoropropylene and tetrafluoroethylene, FKM) with the chemical formula $[-CH_2CF_2-CF_2-CF(CF_3)-CF_2CF_2-]_n$ with four samples for each set have been investigated to study the evolution of frictional properties upon $\gamma$-ray irradiation . The virgin samples were prepared by molding and were then irradiated with $\gamma$-rays at different dose rates but maintaining the total dosage as $10^3$ Gy. Three different dose rates of 14 Gy/h, 61.3 Gy/h and 110 Gy/h were selected for the present study. Three samples from both the sets were irradiated with these dose rates. We carried out the present investigation using scanning probe microscope from NT-MDT, Russia[20]. The AFM and FFM measurements were carrried out in contact mode with the stiffness constant of the cantilever around $\sim$ 0.1 $N/m$ having radius of curvature of the tip $\sim$ 100 Å. The normal force applied for the topographic and frictional (torque) measurement was around $\sim$ 25 nN. All measurements were done ex-situ at room temperature and at ambient condition. The scan size for all the samples were 30 $\mu m$ x 30 $\mu m$.

## III. RESULTS AND DISCUSSION

In frictional force microscopic technique, one can measure simultaneously both the topography and the frictional property by recording the current signals $I_{ver}$ proportional to normal bending of the cantilever (which carries the AFM tip) and $I_{tor}$ which is proportional to torsional bending of the cantilever simultaneously. The current signals $I_{ver}$ represents the topographical height distribution and the current signals $I_{tor}$ represents the frictional component of the sample surface in that region[21–24]. This is described schematically in fig. 1. In fig. 2 we show schematically the data analysis scheme adopted in this investigation. The current signal $I_{tor}$ due to the torsional bending of the cantilever during a line scan along forward (left to right) and backward (right to left) direction is shown. Since the torsional angle ($\beta$) changes sign when moving in the opposite direction the current signal $I_{tor}$ also changes sign. We have also shown in fig. 2 a histogram of the torsional bending of the cantilever proportional to the current $I_{tor}$. We would like to mention that throughout our experiment on all the samples the position of the laser beam on the cantilever was kept on the same spot and the scan speeds were also kept the same. This is very important since we know that variation of the position of the laser beam on the cantilever determines the amplitude of bending of the cantilever and we have also observed that the speed of the cantilever affects the measurement of its bending and frictional property. This aspect of velocity dependent frictional property is under progress with simpler sample surfaces.

In fig. 3 we show the topographical AFM images of EPDM samples: (a) being that of the virgin sample and images (b) to (d) are that of samples with increasing dose rate. The roughness values are tabulated in table I. In fig. 3(a) we can clearly see the topographical changes as a function of irradiation dose rate. The virgin sample (fig. 3(a)) shows agglomeration of grains (here by agglomerate we mean collection of grains and the agglomerate is marked by a circle in fig. 3). With the lowest dose rate, the grain sizes are seen to increase and the agglomerate size decreases (fig. 3(b)). For the next higher dose rate (fig. 3(c)), the grain sizes are smaller than in fig. 3(b) but the agglomerate size seems to decrease slightly with respect to the virgin sample. With the highest dose rate (fig. 3(d)) the grains are more uniformly spread and the agglomerate remains big as in the virgin sample (fig. 3(a)).

In fig. 4 we show the topographical AFM images of the FKM samples: (a) being that of the virgin sample and images (b) to (d) are that of samples with increasing dose rate. From the AFM image we observe that the virgin FKM sample (fig. 4(a)) is smoother than the virgin EPDM (fig. 3(a)) sample, as also indicated by the value of the rms roughness tabulated in table I. For the low irradiation rate we can see blisters appearing (fig. 4(b) marked by circle). For the next higher dose rate the blisters formed are smaller in size (fig. 4(c)) compared to that in fig. 4(b). For the highest dose rate (fig. 4(d)) the blisters formed are very fine and the sample almost appears like the virgin sample. This is reflected in the sample roughness which decreases as a function of the dose rate as shown in table I. We would like to point out here that since the experiments were performed on different samples and it is not an in-situ measurement, the variation in the rms roughness were found to be $\leq 10\%$. The topographical images gives only a qualitative picture of the surface modifications upon any radiation. To detect the process occuring at molecular level, we have carried out FFM measurements on these samples. In the next section we would like to

address the questions:

(1) What happens to the frictional properties of these elastomers upon various $\gamma$-ray dose rate?

(2) Does the frictional property decrease with the decrease in the roughness due to irradiation?

In fig. 5 we show the FFM images of the EPDM samples. We show the FFM images taken during both the forward (i.e., left to right) and the reverse (i.e., right to left) scan direction. Fig. 5(a and b) are for the virgin sample for both the scan directions respectively and fig. 5((c and d),(e and f) and (g and h)) are for the irradiated samples with increasing dose rate for both the scan directions (see figure caption). We observed contrast inversion when the scanning direction is reversed as explained schematically in fig. 2. For the virgin sample homogeneous distribution of fine granular regions can be seen as dark and bright regions in fig. 5(a and b). On irradiation with the lowest dose rate, large granular regions are observed (fig. 5(c and d)). For the next higher dose rate (fig. 5(e and f)) the granular regions appears to be smaller than the previous dose rate and for the highest dose rate (fig. 5(g and h)) the granular regions are as fine and homogeneous as in the virgin sample. It appears that the frictional property upon irradiation with the highest dose rate approaches that of the virgin sample. The FFM images of the FKM samples are shown in fig. 6. We can clearly see the homogeneous spread of the granular regions for the virgin sample (fig. 6(a and b)). On irradiation, we observe similar behaviour as seen for the EPDM samples as a function of dose rate . This scenario can be understood qualitatively - as the dose rate increases, the density of photons per unit time per unit area incident on the sample surface increases. With the increase in dose rate the formation of grains and blisters due to irradiation thus get more and more confined because of higher dose rate and we observe similar results from the AFM images.

One can now quantitatively obtain the overall change in the frictional property by plotting a histogram for both the scan directions as described schematically in fig. 2. In fig. 7(a and b), we have plotted this histogram i.e., number of points (N) vs torsional bending (proportional to $I_{tor}$) obtained from the FFM images for the EPDM samples and the FKM samples respectively. Both the sample set shows a maximum in the histogram for scan obtained in both the scan directions. The average value $I_{tor,avg}$ defined as

$$I_{tor,avg} = \frac{I_{tor,max}(+) + \mid I_{tor,max}(-) \mid}{2} \qquad (1)$$

where $I_{tor,max}(+)$ and $I_{tor,max}(-)$ are the value of $I_{tor}$ each corresponding to the maximum N for the forward and reverse scan directions respectively and are tabulated in table I for both the sets. We observe from fig. 7(a) that the virgin EPDM sample shows a very low frictional coefficient compared to the irradiated samples since the maximum of the histogram has the lowest torsional bending (see curve (a)) for this sample. On irradiation with the lowest dose rate the maximum of the histogram shifts to higher torsional bending ($I_{tor}$) indicating an increase in the overall frictional coefficient (see curve (b)). On further increase in the dose rate we observe that the maximum shifts back to lower value of $I_{tor}$ indicating a decrease in the overall frictional coefficient of the sample (see curve (c)). With still further increase in dose rate the frictional coefficient approaches a stable value. Thus we observe that the overall frictional coefficient of the sample as a function of dose rate initially increases and then decreases and stablises to a lower value with increasing dose rate. For the FKM samples the histogram shown in fig. 7(b) shows a similar behaviour. At a closer look we observe an additional interesting feature i.e., a very sharp peak at very low values of torsional bending for the FKM samples (inset of fig 7(b)). Additionally the broad maxima becomes a shoulder to the low value peak in the histogram as the dose rate increases. These shoulders are marked by arrows in fig. 7(b). In table I, we have tabulated the broad maxima and the shoulder values of $I_{tor,avg}$ (phase I). The two maxima in the histogram signifies the existence of two distinct phases, with two different frictional coefficients. The appearance of the sharp peak on irradiation at lower value of $I_{tor} < 0.5$nA (see table I, phase II) arises due to the formation of another phase having lower frictional coefficient. With increasing dose rate the height of the sharp peak increases indicating that the formation of second phase is caused by irradiation and the quantity of this phase increases with increase in dose rate. By adopting the histogram scheme we obtain two main results:

(1) The overall frictional coefficient of the sample as a function of dose rate initially increases and then decreases and stablises to a lower value with increasing dose rate.

(2) Two maxima was observed in the histogram for the FKM sample indicating generation of two distinct phases having two different frictional coefficients whereas the EPDM sample shows only one maximum.

Below we will discuss the result (2) first and then the result (1) for clarity.

The observed features i.e., double maxima and the single maximum in the histogram can be explained as follows. The FKM sample is a copolymer consisting of three different types of monomers (1) Hexa fluoro propylene ($CF_2=CF-CF_3$) and (2) tetrafluoroethylene ($CF_2=CF_2$) having only a C-F bond and (3) Vinylediene Fluoride ($CH_2=CF_2$) having both C-H and C-F bonds. FKM with two dissimilar bonds gets phase seperated on irradiation giving rise to chemical or structural changes hence leading to two distinct phases having different distinct overall frictional coefficients. Further spectroscopic study is needed to identify the two phases. On the other hand, the EPDM sample is composed of ethylene ($CH_2=CH_2$) and propylene ($CH_2=CH-CH_3$) both having only C-H bonding and hence cannot show any phase seperation. Thus, it shows only one component of frictional coefficient on irradiation. This comparison shows that the analysis using histogram scheme reveals whether there exists different phases of materials with different frictional coefficients in the sample.

The other important result obtained from this investigation was that, both the samples exhibit overall increase of frictional coefficient on lower dose rate and on further increase of dose rate the frictional coefficient decreases. The change in the frictional coefficient ($\Delta\mu$) is proportional to the difference in $I_{tor,avg}$ value for each of the irradiated samples and the value of $I_{tor,avg}$ of the virgin sample. Normalizing $\Delta\mu$ with the value of $I_{tor,avg}$ of the virgin sample we obtain the normalized change in the overall frictional coefficient $\Delta\mu/\mu$ as,

$$\Delta\mu/\mu = \frac{(I_{tor,avg})_{irr} - (I_{tor,avg})_{vir}}{(I_{tor,avg})_{vir}} \qquad (2)$$

where the subscript *irr* and *vir* represents irradiated and virgin samples respectively. The calculated values of $\Delta\mu/\mu$ for both the sets are tabulated in table I. We observe that the overall frictional coefficient ($\Delta\mu/\mu$) increases by more than 4 times for the EPDM sample and more than 20 times for the FKM sample for $\gamma$-ray dose rate of 14 Gy/h and on further increase in dose rate, the normalized frictional coefficient decreases in both the samples. The second phase formed on irradiation for the FKM sample does not show much change in the overall frictional coefficient. The initial increase of the frictional coefficient observed in case of both the samples upon $\gamma$-ray irradiation can be attributed to formation of free radicals due to chain scission. Formation of free radicals beyond a critical concentration leads to crosslinking among the macromolecular radicals at higher dose rate. Formation of polymer chains due to the recombination of free radicals (crosslinking) lead to lowering of frictional coefficient. The tip of the AFM adheres to these free radicals present on the sample surface and with the onset of crosslinking the frictional coefficient decreases. Hence, one can conclude from the above experimental results that when the elastomers are exposed with higher dose rate then the probability of chain scission is larger and the macroradicals formed in close proximity promotes the crosslinking of the polymers. Low concentration of free radicals will not promote crosslinking. The local frictional property of these materials depends on the ratio of crosslinking to chain scission. The frictional coefficient decreases with increase in degree of crosslinking. Thus with FFM, one can detect the competitive behaviour between the chain scission and the crosslinking process.

## IV. CONCLUSION

Frictional force microscopic technique has been used to detect onset of chain scission and crosslinking of $\gamma$-ray irradiated elastomer samples. In the present study we observe initial increase of the frictional coefficient upon $\gamma$-ray irradiation with low dose rate and the frictional coefficient decreases upon irradiation with higher dose rate for both the EPDM and the FKM elastomer samples while maintaining the total dose constant. The increase in the frictional coefficient has been attributed to formation of free radicals and the decrease in frictional coefficient has been attributed to onset of crosslinking. The crosslinking takes place beyond certain critical concentration of free radicals produced. With the FFM technique we could distinguish the presence of coexisting phases with different frictional coefficients. We could infer from the reduction of the surface roughness upon $\gamma$-ray irradiation that the frictional property of the sample determined using FFM depends on the atomic and molecular interaction between the tip and the sample surface rather than on the asperities (roughness) of the sample surface. The radical formation and crosslinking occurs at molecular level, thus FFM analysis can help in understanding these processes at the molecular level better than macroscopic tribological techniques.

**Acknowledgement**
We would like to thank Dr. R. K. Singh, Additional Director, DMSRDE, Kanpur for providing us the moulded elastomer samples used for the present studies.

# References

[1] H. Wilski, Radiat. Phys. Chem., **29** 1 (1987)

[2] C. B. Saunders, Radiation processing in the plastic industry: current commercial applications, Report AECL 9569 (1988)

[3] J. Davenas, I. Stevenson, N. Celette, G. Vigier and L. David, Nucl. Instr. and Meth. B, **208** 461 (2003)

[4] I. Banik and A. Bhowmick, Radiat. Phys. Chem., **58** 293 (2000)

[5] A. A. Basfar, M. M. Abdel-Aziz and S. Mofti, Polym. Degrad. Stab., **66** 191 (1999)

[6] M. M. Abdel-Aziz and A. A. Basfar, Polymer Testing, **19** 591 (2000)

[7] I. Banik and A. Bhowmick, Radiat. Phys. Chem., **54** 135 (1999)

[8] M. D. Chipara, V. V. Grecu, M. I. Chipara, C. Ponta and J. R. Romero, Nucl. Instr. and Meth. B, **151** 444 (1999)

[9] P. S. Majumder and A. K. Bhowmick, Wear, **221** 15 (1998)

[10] P. Palmas, R. Colsenet, L. Lemarie and M. Sebban, Polymer, **44** 4889 (2003)

[11] A. Oshima, S. Ikeda, E. Katoh and Y Tabata, Radiat. Phys. Chem., **62** 39 (2001)

[12] M. F. Vallat, F. Ruch and M. O. David, Eur. Polym. J., **40** 1575 (2004)

[13] T. Seguchi, Radiat. Phys. Chem., **57** 367 (2000)

[14] A. Rivaton, S. Cambon and J. -L. Gardette, Nucl. Instr. and Meth. B, **227** 357 (2005)

[15] A. Rivaton, S. Cambon and J. -L. Gardette, Nucl. Instr. and Meth. B, **227** 343 (2005)

[16] T. Zaharescu, V. Meltzer and R. Vilcu, Polym. Degrad. Stab., **61** 383(1998)

[17] T. Zaharescu, S. Jipa, M. Giurginca and C. Podina, Polym. Degrad. Stab., **62** 569 (1998)

[18] T. Zaharescu, E. Feraru and C. Podina, Polym. Degrad. Stab., **87** 11 (2005)

[19] I. Banik, A. K. Bhowmick, S. V. Raghavan, A. B. Majali and V. K. Tikku, Polym. Degrad. Stab., **63** 11 (1999)

[20] NT-MDT See:http://www.ntmdt.ru

[21] R. G. Cain, M. G. Reistma, S. Biggs and N. W. Page, Rev. Sci. Instrum. **72**8 3304 (2001)

[22] J. E. Sader, Rev. Sci. Instrum. **74**4 2438 (2003)

[23] J. E. Sader and C. P. Green, Rev. Sci. Instrum. **75**4 878 (2004)

[24] P. Bilas, L. Romana, B. Kraus, Y. Bercion and J. L. Mansot, Rev. Sci. Instrum. **75**2 415 (2004)

Figure Captions

**Fig.1** Schematic diagram of the vertical and torsional bending of the cantilever due to tip-surface interaction. The photodiode shows four quadrant A, B, C and D. The reflected laser light from the cantilever have vertical motion on photodiode due to vertical bending and lateral motion due to torsional bending of the cantilever.

**Fig.2** Schematic illustration of $I_{tor}$ due to the torsional bending of the cantilever for the forward and reverse direction of scan showing sign reversal of $I_{tor}$. For the respective scan direction we have plotted histogram of $I_{tor}$ vs. N (N is number of points per $I_{tor}$). [Note: the maxima of $I_{tor}$ becomes minima on reversal of scan direction leading to contrast inversion in FFM images.]

**Fig.3** The topography images of EPDM samples (a) virgin sample and with increase in dose rates (b) 14 Gy/h, (c) 61.3 Gy/h and (d) 110 Gy/h.

**Fig.4** The topography images of FKM samples (a) virgin sample and with increase in dose rates (b) 14 Gy/h, (c) 61.3 Gy/h and (d) 110 Gy/h.

**Fig.5** FFM images of the virgin (a,b) and irradiated samples with increase in dose rates of (c,d) 14 Gy/h, (e,f) 61.3 Gy/h and (g,h) 110 Gy/h for EPDM samples.

**Fig.6** FFM images of the virgin (a,b) and irradiated samples with increase in dose rates of (c,d) 14 Gy/h, (e,f) 61.3 Gy/h and (g,h) 110 Gy/h for FKM samples.

**Fig.7** (a) Histogram showing number of points per $I_{tor}$, N vs. $I_{tor}$ for EPDM sample, curves a, b, c and d are for virgin, 14 Gy/h, 61.3 Gy/h and 110 Gy/h irradiated sample respectively (b) Similar histogram for FKM sample and arrows indicates the shoulder as described in the text.

TABLE I: Roughness obtained from AFM images, average current signals corresponding to trosional bending of the cantilever at maximum N or shoulders for FKM sample obtained from fig. 7(a and b) for phase I in EPDM and FKM samples and phase II for the FKM samples and $\Delta\mu/\mu$ for each of the phases of EPDM and FKM samples.

| Sample | Roughness (nm) ±10% | $I_{tor,avg}$ (nA) Phase I | $I_{tor,avg}$ (nA) Phase II | $\Delta\mu/\mu$ Phase I | $\Delta\mu/\mu$ Phase II |
|---|---|---|---|---|---|
| EPDM (virgin) | 150 | 0.2 | - | - | - |
| EPDM (14 Gy/h) | 110 | 1.05 | - | 4.3 | - |
| EPDM (61.3 Gy/h) | 105 | 0.43 | - | 1.2 | - |
| EPDM (110 Gy/h) | 100 | 0.47 | - | 1.3 | - |
| FKM (virgin) | 90 | 0.27 | 0.10 | - | - |
| FKM (14 Gy/h) | 75 | 6.4 | 0.10 | 22.7 | - |
| FKM (61.3 Gy/h) | 65 | 2.8 | 0.24 | 9.4 | 1.4 |
| FKM (110 Gy/h) | 55 | 1.3 | 0.28 | 3.8 | 1.8 |

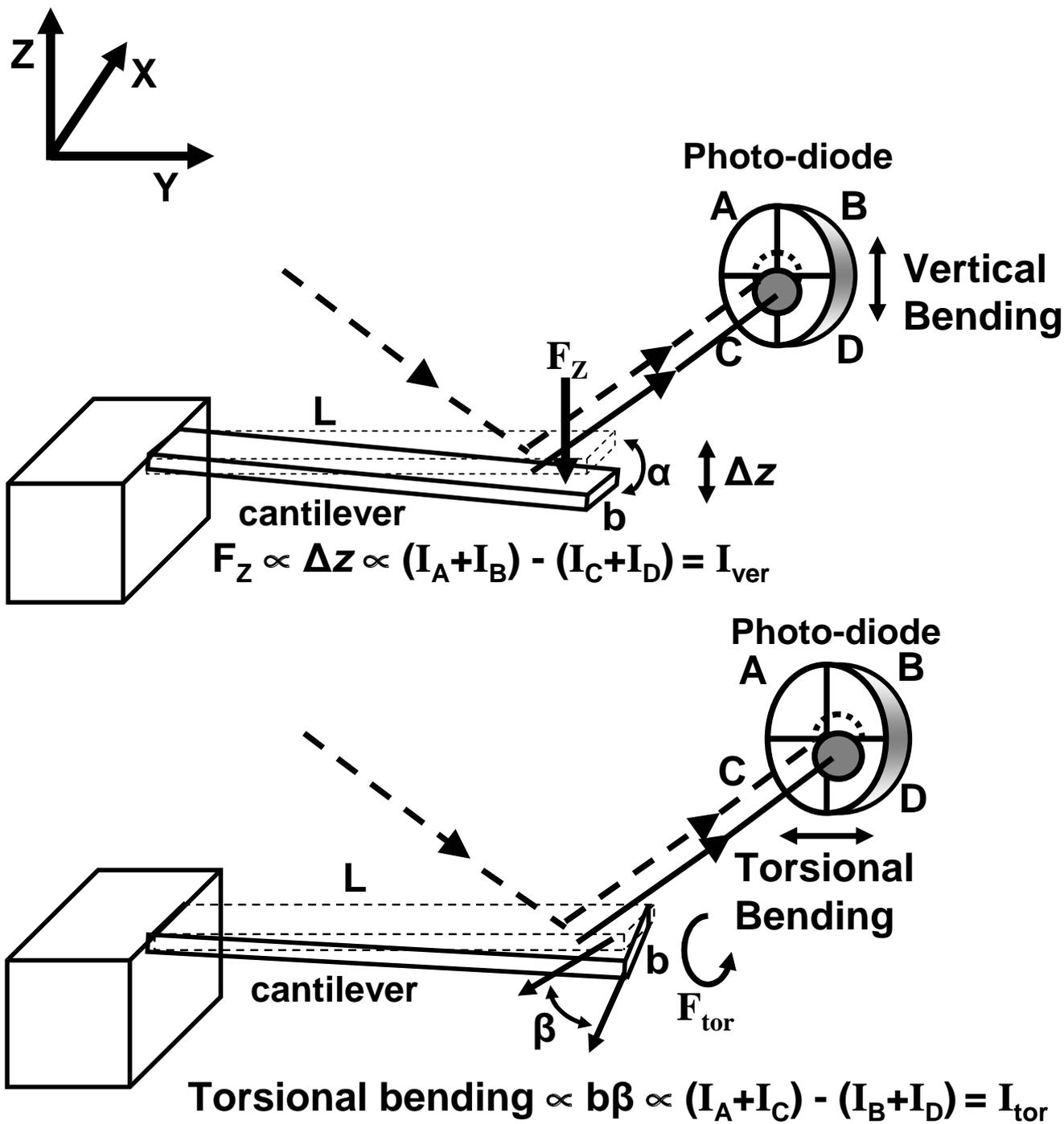

Fig 1. S. Banerjee et.al.

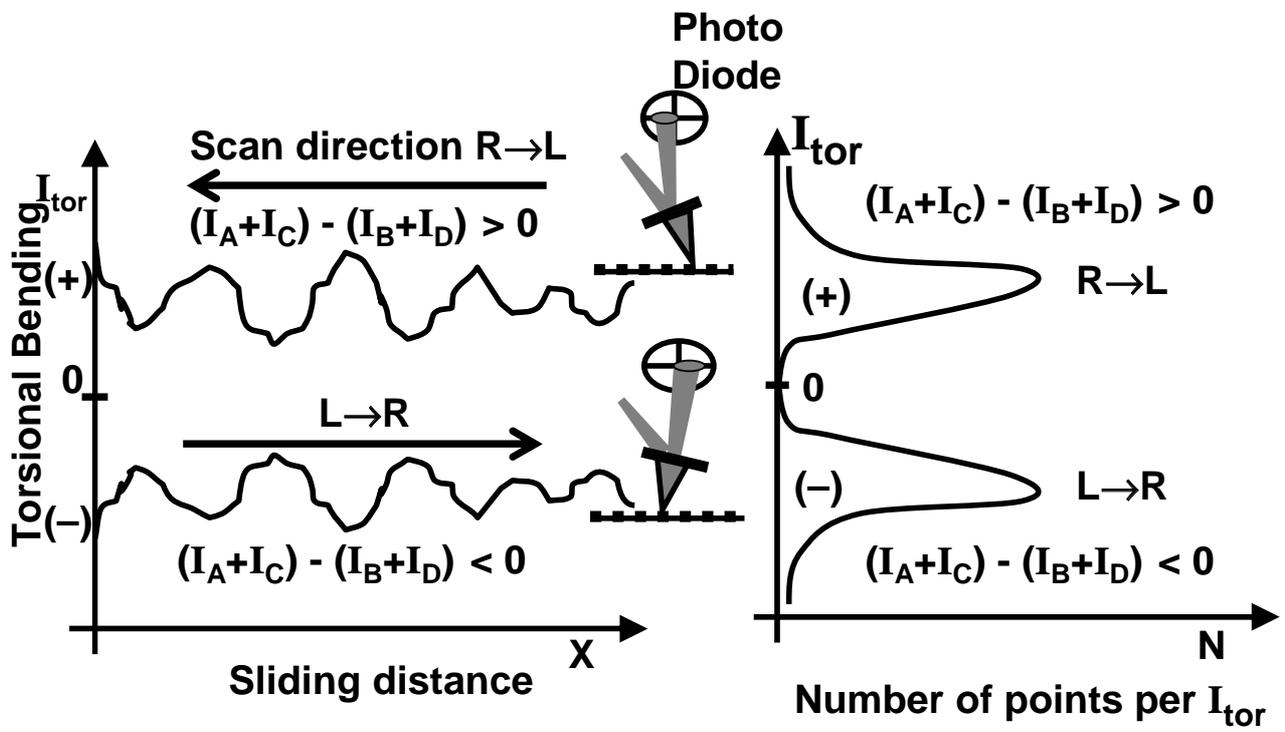

Fig 2. S. Banerjee et.al.

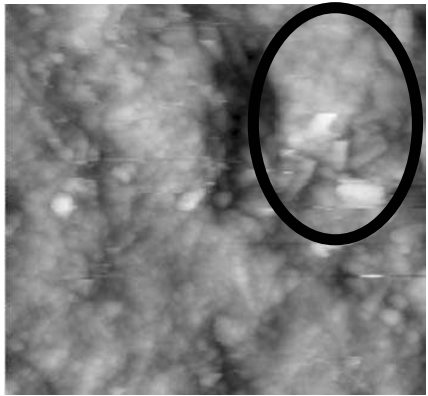 (a)

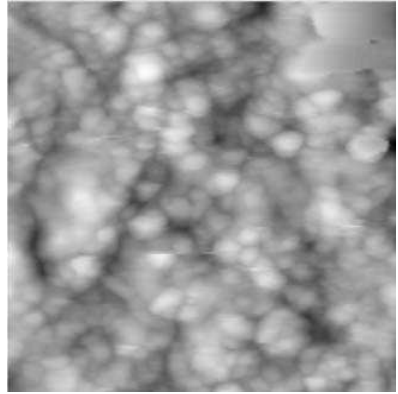 (b)

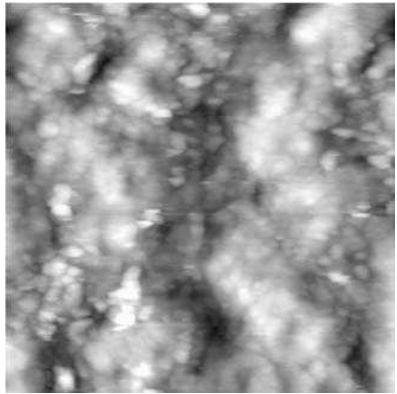 (c)

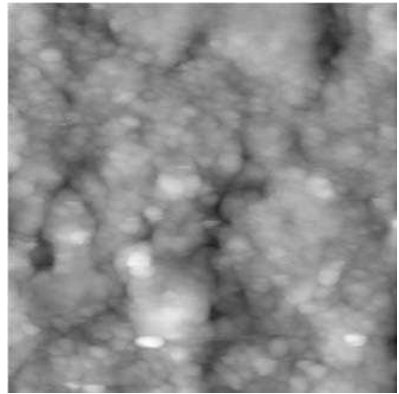 (d)

Fig 3. S. Banerjee et.al.

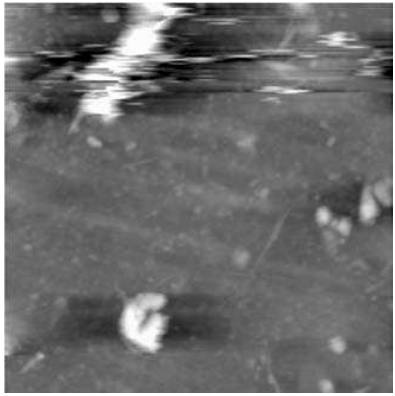 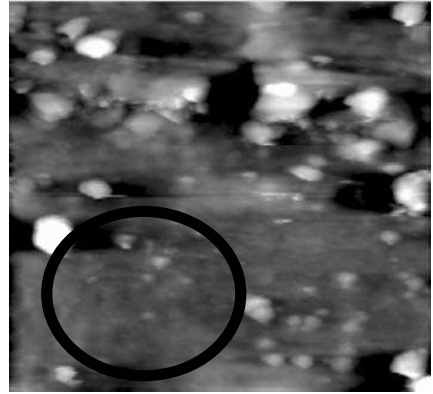

**(a)** **(b)**

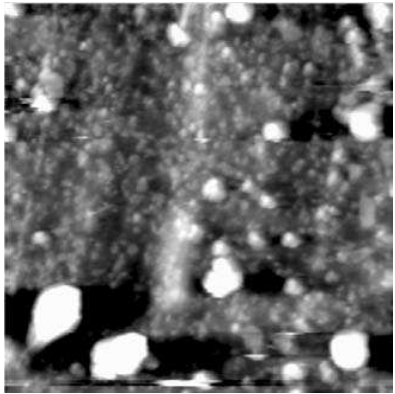 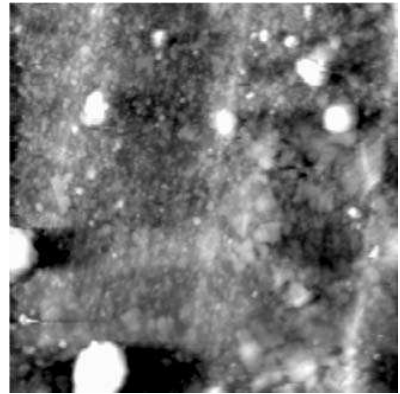

**(c)** **(d)**

Fig 4. S. Banerjee et.al.

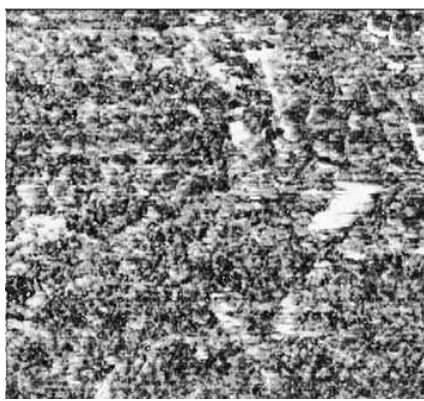 (a)
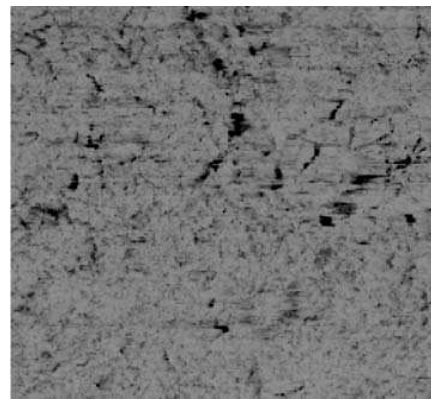 (b)
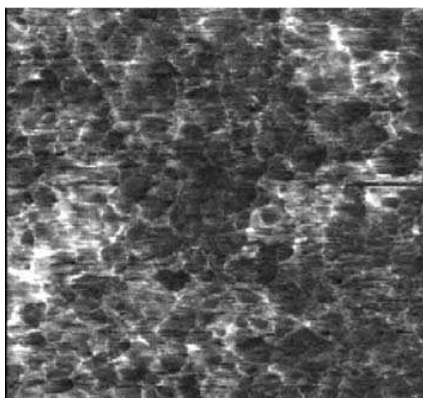 (c)
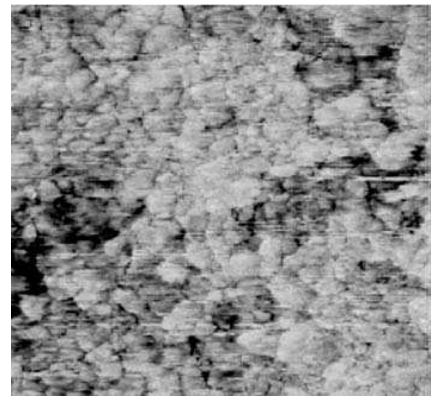 (d)
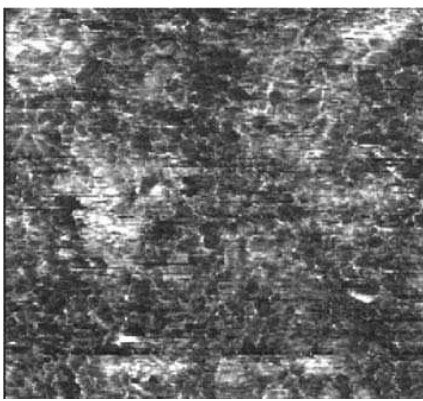 (e)
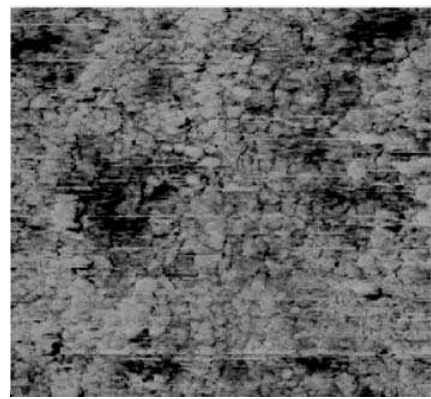 (f)
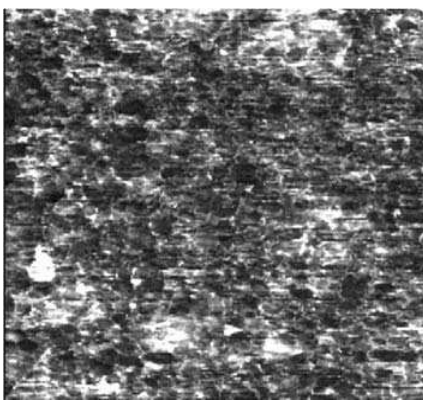 (g)
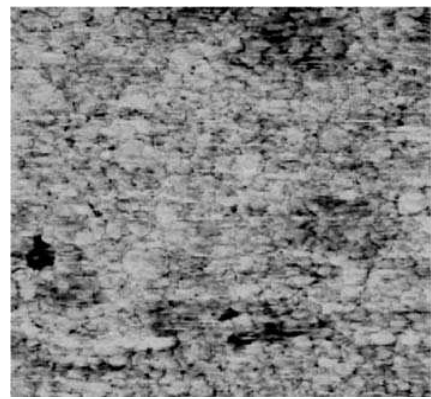 (h)

Fig 5. S. Banerjee et.al.

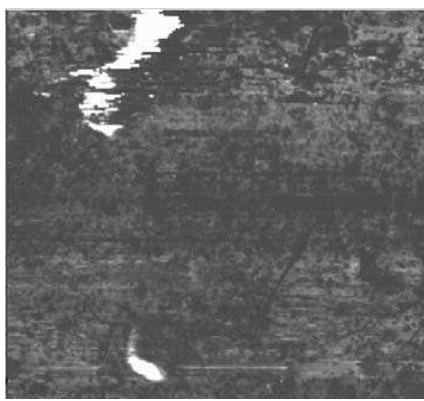 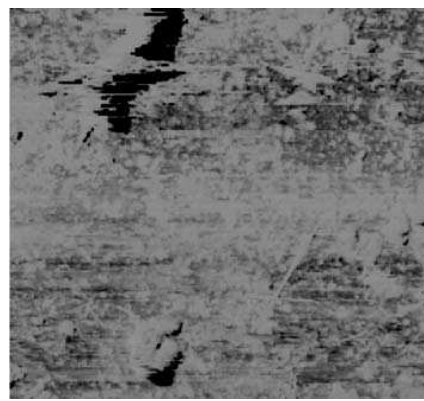
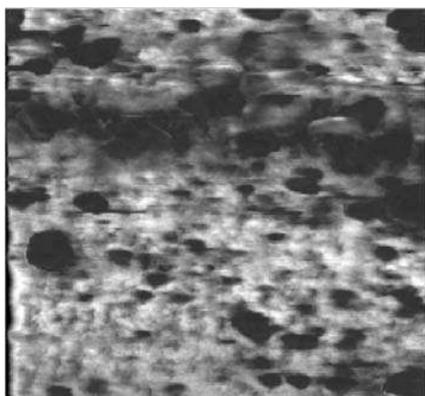 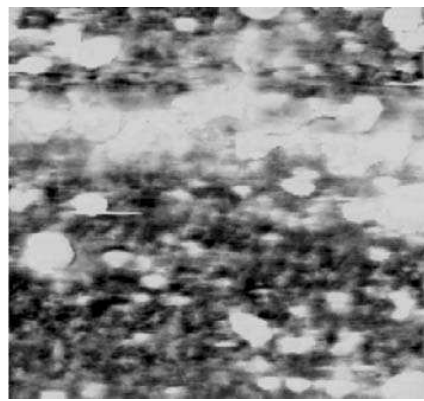
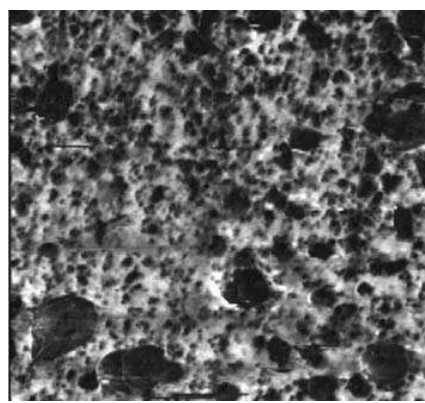 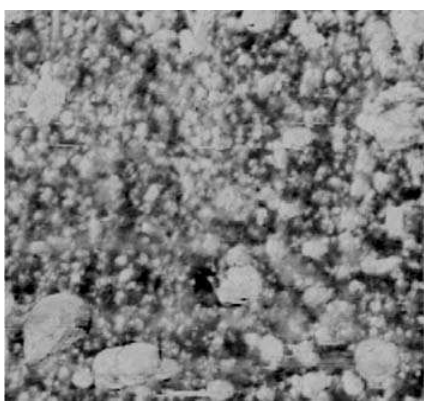
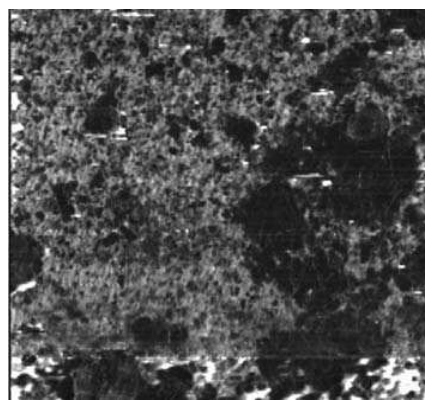 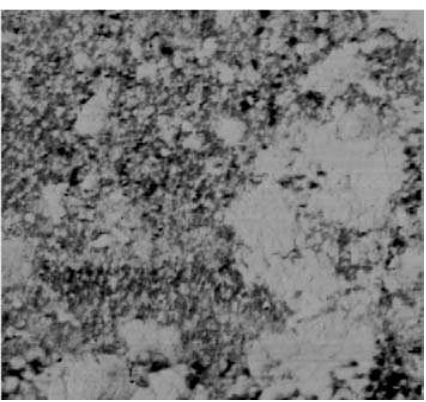

Fig 6. S. Banerjee et.al.

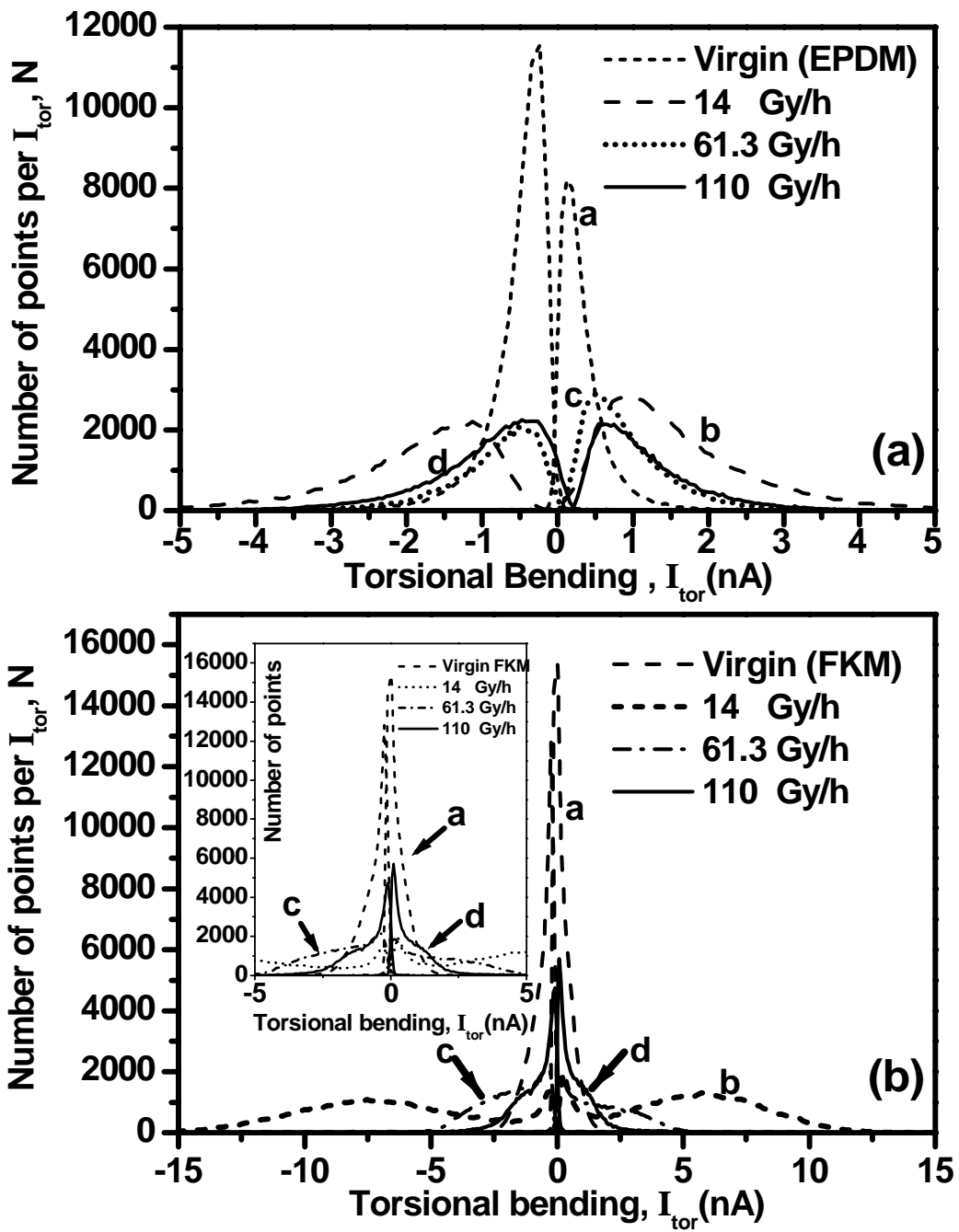

Fig 7. S. Banerjee et.al.